%% file: main.tex
\RequirePackage{fix-cm}
\documentclass{svjour3}                     % onecolumn (standard format)
\smartqed  % flush right qed marks, e.g. at end of proof
\usepackage{graphicx}
\usepackage{xspace}

\usepackage{color}
\usepackage{soul}
\usepackage{natbib}
\usepackage{verbatim}
\renewcommand{\cite}{\citep}
\usepackage{booktabs} %for Top/bottom rules on tables

\makeatletter
\AtBeginDocument{%
  \def\doi#1{\url{https://doi.org/#1}}}
\makeatother
\usepackage{underscore}

\newcommand{\fig}[4][0.9]{
\begin{figure}
  \centering
  \includegraphics[width=#1\textwidth]{#2}
  \caption{#4}
  \label{fig:#3}
\end{figure}
}

\newcommand{\qt}[2]{%
\textit{``#2''} {\textsc{[#1]}}\xspace%
}

\newcommand{\qtn}[1] {\xspace\textsc{[#1]}\xspace}

\makeatletter
\newcount\my@repeat@count
\newcommand{\myrepeat}[2]{%
  \begingroup
  \my@repeat@count=\z@
  \@whilenum\my@repeat@count<#1\do{#2\advance\my@repeat@count\@ne}%
  \endgroup
}
\makeatother
\usepackage{wasysym}

\definecolor{myRed}{RGB}{238, 107, 115}
\definecolor{myOrange}{RGB}{250, 199, 16}

\newcommand{\numUses}[1]{
\vspace{-5pt}
\color{myRed}{
\scriptsize
\myrepeat{#1}{\CIRCLE\hspace{0.001cm}}
}\color{black}
\newline
}

\newcommand{\numUsesN}[1]{
\vspace{-5pt}
\color{myOrange}{
\scriptsize
\myrepeat{#1}{\CIRCLE\hspace{0.001cm}}
}\color{black}
}

\newcommand{\numUseAlign}[0]{
\color{white}{
\scriptsize
\CIRCLE
\newline
}\color{black}
}

\begin{document}

\title{Grasping AI: experiential exercises for designers%\thanks{Grants or other notes
}


\author{Dave Murray-Rust \and
        Maria Luce Lupetti \and Iohanna Nicenboim \and Wouter van der Hoog
}


\institute{Dave Murray-Rust \at
              Human Centred Design\\
              Industrial Design Engineering \\
              TU Delft, Netherlands \\
              \email{d.s.murray-rust@tudelft.nl}       %  \\
           \and
           Maria Luce Lupetti \at \email{m.l.lupetti@tudelft.nl}
           \and
           Iohanna Nicenboim \at \email{I.Nicenboim@tudelft.nl}
           \and
           Wouter van der Hoog \at \email{wvanderhoog@gmail.com}
}

\date{Received: date / Accepted: date}

\maketitle

\begin{abstract}
Artificial intelligence (AI) and machine learning (ML)  are increasingly integrated into the functioning of physical and digital products, creating unprecedented opportunities for interaction and functionality. However, there is a challenge for designers to ideate within this creative landscape, balancing the possibilities of technology with human interactional concerns.
We investigate techniques for  exploring and reflecting on the interactional affordances, the unique relational possibilities, and the wider social implications of AI systems. We introduced into an interaction design course (n=100) nine  ‘AI  exercises’  that draw on more than human design, responsible AI, and speculative enactment to create experiential engagements around AI interaction design.
We find that exercises around metaphors and enactments make questions of training and learning, privacy and consent, autonomy and agency more tangible, and thereby help students  be more reflective and responsible on how to design with AI and its complex properties  in both their design process and outcomes. 

\keywords{Artificial Intelligence \and design \and prototyping \and design education \and experiential methods \and AI exercises \and more-than-human design}
\end{abstract}

\input{01-Introduction.tex}
\input{02-Background.tex}
\input{03-Materials.tex}

\input{04-Findings.tex}

\input{05-Discussion.tex}

%\begin{acknowledgements}
%\end{acknowledgements}

\section{Acknowledgements}
We would like to thank our course collaborators for assistance in both running and analysing the course, in particular Ianus Keller, Aadjan van der Helm, Tomasz Jaskeiwicz, Dieter Vandoren, Gijs Huisman, Nazli Cila and Martin Havranek, as well as Seowoo Nam for graphic design and data collection around the methods. Thanks to the others in the Human Centred Design department who helped us to think about and frame this work, in particular Elisa Giaccardi. This work was partly supported by the Microsoft Research PhD fellowship awarded to Iohanna Nicenboim. Finally, as an piece of educational research we would like to thank the students on the course for their hard work, thoughtfulness, creativity and boldness.

\section{Additional Information}
On behalf of all authors, the corresponding author states that there is no conflict of interest.
The datasets generated during and/or analysed during the current study are available from the corresponding author on reasonable request.

\bibliographystyle{spbasic}      % basic style, author-year citations
\bibliography{main}   % name your BibTeX data base

\end{document}

%% file: 01-Introduction.tex
\section{Introduction}

Designers increasingly need to develop a facility with artificial intelligence, as it becomes part of the way that products and services function and appears in an increasing number of the contexts in which designers work \cite{benjamin2021MachineLearning, dove2017UXDesign}. However, there are several challenges for design students in engaging with AI, from the broadness of the term AI and the fuzziness with which it is applied \cite{littman2021GatheringStrength}, to the difficulty of getting to grips with the technical and computational complexities of these systems \cite{yang2020ReexaminingWhether,nicenboim2022ExplanationsShared}. These challenges around understanding and making sense of the new capabilities of AI become urgent as the technology emerges from it's latest winter into a new spring, developing at a fast pace \cite{littman2021GatheringStrength, samoli2020AIWatch, floridi2020AIIts}. 

The range of techniques for making creative use of AI has been rapidly growing: Runway offered easy access to generative spaces and now video \cite{SystemRunway}; EdgeImpulse offers sound and gesture classification for microcontrollers with training through a web interface \cite{SystemEdgeImpulse}; the current sets of generative image models such as DALL-E, Midjourney and StableDiffusion %\cite{SystemDALL-E2, SystemMidjourney, SystemStableDiffusion} 
and language models (ChatGPT etc.) %\cite{SystemGPT-3,SystemGPT-J} 
allow a natural language interaction through the use of prompts. Along with learning materials for more traditional toolkits \cite{SystemTensorFlow, SystemOpenCV} and model development and exchange initiatives \cite[e.g.][]{SystemHuggingFace} these form a downward pressure on the technical barrier to entry, even as the complexity of the underlying models increases. The conceptual barrier can remain high, though, reducing the possibility for designerly engagement and appropriation. There is a large jump from ``my first ML model'' to understanding the implications of ML technology, and designers often want – and need – to engage with these implications. Creating models in practice helps, but this needs conceptual framing to help direct and contextualise the activities - for example, in related fields, courses such as Creative Applications of Deep Learning \cite{mital2016CreativeApplications} and the more provocative follow up Cultural Appropriation with Deep Learning \cite{mital2021CulturalAppropriation} look at visual practice, or Machine Learning for Musicians and Artists \cite{fiebrinkMachineLearning} unpack these systems for creative practitioners.

However, simply thinking about why it is hard for designers to appropriate AI technologies into their practices also misses a key question: what can design practice bring to the development and understanding of AI systems \cite{benjamin2021MachineLearning}, especially as the technologies become more pervasive and more collaborative \cite{wang2020HumanhumanCollaboration}.  
Designers have their own strategies for making use of, critiquing and appropriating new technologies \cite{westerlund2017DealingWicked}, so there is an interest in understanding what designerly methods could reveal about human AI relations, particularly where it involves interactions between humans and technological systems - considering the “social, political, ethical, cultural, and environmental factors of implementing AI into daily human-to-computer interactions” \cite{wong2018DesignFiction}. Design research methods, speculations \cite{auger2013SpeculativeDesign, kirman2022ThinkingOutside}, fictioning \cite{forlano2014DesignFiction,wong2017RealfictionalEntanglements, troiano2021AreWe, benjamin2023EntopticField}, probes and toolkits \cite{sanders2014ProbesToolkits}, more than human design \cite{coulton2019MoreThanHuman} and the general practices of Research through Design (RtD) \cite{giaccardi2019HistoriesFutures, stappers2017ResearchDesign}, are all well suited to thinking into the socio-technical aspects \cite{holton2021WhereAre, sartori2022SociotechnicalPerspectivea, theodorou2020EthicalSociolegal}, possibilities, and implication of AI in everyday life, just as they have been applied to understanding digital sensing technologies \cite{pierce2021EccentricSensing}, blockchains \cite{murray-rust2022BlockchainUnderstanding}, the future of automation \cite{cavalcantesiebert2022MeaningfulHuman} and so on.

%\todo{Check!}
Our aim is to help students to design products and services that make use of AI technologies, while developing a critical understanding of its implications.  This means articulating both the technical and relational aspects of AI so that meaningfully shape the development of products, services and systems even if they are not intimately familiar with the technical details of its operation. As such we are looking for ways to sensitize interaction designers to AI, to create experiences rather than explanations. In relation to the typology developed by \citet{yang2020ReexaminingWhether}  of ways to aid designers around AI, our work contributes to the early stages of `creating AI-specific design processes’ by probing with concrete exercises ways in which educators can support designers in ideating in a space mediated by the capabilities and implications of AI systems.

In order to explore this space, we created a set of methods for designing AI driven products and services (Section \ref{sec:methods}) that draw on theories about how people relate to technology and AI in particular. These methods take the form of short, autonomous, experiential exercises that can be used to develop and enrich the design of interactive technological products and services. We introduced these exercises partway through an interaction design course (Section \ref{sec:course-context}), where, students (n=100) in small groups (n=28) are asked to design future products and services through iterative prototyping and testing. We collected an immediate written reaction from each group as to what the students had done with the materials, and the aspects they found useful or resonant. We interviewed a self-selecting subset of the students (n=12) and their coaches (n=7) to dig deeper into questions of how the methods had changed their understandings and relations to AI.

To explore the potential of these methods, we investigate the following research questions: 
\begin{itemize}
\item \textbf{RQ1:} How do the exercises stimulate and modulate changes to the students’ design process to accommodate AI, in particular the way that they are conceptualising and prototyping their projects?
\item \textbf{RQ2:} How do these experiential exercises affect  student’s grasp of AI and ML, in  particular in relation to  interactional, relational and contextual qualities which are key points in the recent theoretical developments in AI within HCI?
\item \textbf{RQ3:} How do the exercises help to develop a critical design perspective while engaging with AI technology as a socio-technical system?
\end{itemize}

Through investigating and discussing these research questions, the contributions of this work are: 
\begin{enumerate}
\item A set of exercises that translate theoretical developments in design and AI, into experiential exercises for designers that can be carried out autonomously, with reflection on the experiential, pragmatic and reflective qualities that made the exercises effective. These exercises are available at [redacted] for future use and development.
\item Insights into how and for what to apply these exercises in a pedagogical context to support design processes for creating AI enabled products and services.
\item Insights about how these exercises affected student's reasoning and design activities, bringing agency, relationality and criticality alongside development of technical facility.
\item Methodological reflections around the possibilities afforded by the methods and how these contribute to nurturing a uniquely designerly AI culture that supports future design education.
\end{enumerate}

%% file: 02-Background.tex
\section{Background}
Working with AI presents particular challenges for designers. One of them is around engaging with emerging and complex technologies, with different behaviours from traditional design materials. \citet{yang2020ReexaminingWhether} point at two key challenges: the uncertainty about the capability of AI systems, and the complexity of their outputs. The second challenge is around understanding AI, given that the metaphors and imaginaries around it, obscure the real processes that are needed for maintaining such a technology \cite{murray-rust2022MetaphorsDesigners}. Contrary to competing terms like ‘complex information processing systems’ or  ‘machine intelligence,’ the term AI fires the mind with ideas of human-like reasoning. While these imaginaries seem to be better for marketing, they are certainly no good for developing a grounded sense of the capabilities of AI as a technology.  \cite{hildebrandt2020SmartTechnologies} 
\subsection{Designing AI}
Despite the challenges for designers to engage with AI, there are currently many areas where design and AI touch on each other. 

At a low level, there is growing attention to the meeting of AI and user experience (UX), as the new possibilities offered by the technology allow new kinds of interaction, and are susceptible to new pitfalls. Techniques are emerging that help to create user interfaces that work with AI systems \cite{subramonyam2021ProcessModel, subramonyam2021ProtoaiModelinformed}, or support innovating AI-powered services and systems within enterprises \cite{yildirim2022HowExperienced}. This can be seen in Microsoft’s guidelines for human-AI interaction \cite{amershi2019GuidelinesHumanAI}, or Google PAIR’s Guidebook \cite{2020PeopleAI}, as well as efforts to bring HCI together with AI \cite{inkpen2019WhereHuman}. Recently, the identification of `AI capabilities' \cite{yildirim2023CreatingDesign} provides a concrete way to think about design spaces for the interactional aspects of computational system.  
Negotiation between AI and HCI can be deep and subtle: interactional affordances help to calibrate trust and reliance between humans and AI;  conceptual metaphors sculpt the relations formed with conversational agents \cite{jung2022GreatChain,khadpepranav2020ConceptualMetaphors}; and appropriate abstractions make AI qualities at hand for creative practitioners \cite{fiebrink2019MachineLearning, tremblay2021EnablingProgrammatic}.
User experience, in its broader sense, goes beyond designing the immediate experiences, with work starting to consider how to develop frameworks for creating more or less personal, dependent and discretionary interfaces \cite{kliman-silver2020AdaptingUser}, or at how to generate heuristic models of meaningful engagement with AI artworks \cite{hemment2022HeuristicModel}. 

Zooming out slightly, a collection of theoretical issues around AI relate to  emerging fields in the third (and fourth) wave HCI and the  philosophy of design communities. Scholars in those areas have grappled with how the concepts used in design and HCI practices might be tied to the industrial era, and how they might have to change and adapt to the new kinds of products and materials which are enabled by AI. This includes lines of research such as post-industrial design, more than human practices \cite{giaccardi2020TechnologyMoreThanHuman}, entanglement thinking \cite{frauenberger2020EntanglementHCI, murray-rust2019EntangledEthnography, hodder2016StudiesHumanThing}, fluid assemblages and multi-intentionality \cite{redstrom2018ChangingThings,wiltse2020RelatingThings}. All these offer vibrant pictures of a new set of relationships between humans and the material world in which both entities `co-constitute' each other. Along with reorienting the relationships between humans and non-humans, scholars within those fields have been rethinking  what it is to `do design', breaking with traditions focused on the subject-object dichotomy \cite{giaccardi2020TechnologyMoreThanHuman}, where design goes beyond a mere problem solving enterprise and becomes an ongoing and more inclusive practice. 
Although these theoretical developments seem to be gathering momentum, they are still not fully translated into practical tools for designers -- the jump from Barad’s agential realism \cite{barad2007MeetingUniverse} to configurations of bits and programs takes careful work \cite{scurto2021PrototypingMachine, seymour2022RespectLens, sanches2022DiffractioninactionDesignerly}. 

At a broader scale, beyond the immediate interactions, some of the theories and practices at play are oriented towards engineering particular system qualities and properties: value sensitive design can help to make sense of fluid and evolving systems \cite{dereuver2020DigitalPlatforms}, where many different human values may be at play \cite{yurrita2022MultistakeholderValueFramework, fish2021ReflexiveDesign, shen2021ValueCards}; questions of meaningful human control modulate the relations of responsibility between humans and automated systems \cite{cavalcantesiebert2022MeaningfulHuman}, as does responsible AI design \cite{benjamins2019ResponsibleAI}. Here, design is an instrumental part of making systems behave in certain ways.  AI ethics is a broad field \cite{hagendorff2020EthicsAI}, and well as directly affecting system properties, work from the Fairness Accountability and Transparency (FAccT) community looks to support documentation that helps maintain these properties in communities, such as documentation for models and datasets \cite{mitchell2019ModelCards,gebru2018DatasheetsDatasets}, and  the ethical aspects of system development \cite{mohammad2021EthicsSheets,murray-rust2022EthicAmanuensisSupporting}.

\subsection{AI and Design Education}
The specificities and challenges of AI and ML technologies add up to an ongoing discourse at the intersection of design education and technological progress. Traditional formats and scopes for carrying out design are being questioned and revised, with canonical, linear, causal, and instrumental approaches being criticized in favor of novel models inspired by complexity theory, system science, and practical philosophy. This moves towards an aim of reconceptualizing design as a moral act \cite{findeli2001RethinkingDesigna, lin2014RethinkingDesign}. Designers and design researchers, in fact, are increasingly recognized as actors whose decisions have ethical as well as political implications \cite{lloyd2019YouMake}.
Parallel to this, the societal implications of AI and ML are more clearly pervasive and unpredictable. However, when introduced in design education, these technologies are typically either approached as the ultimate tools to learn or used as ‘context’ for grounding alternative and critical design explorations. On the one hand, ML courses are increasingly offered to design students for promoting ML/AI literacy but remain an addition to the main curricula, rather than being integrated into project courses (as in \citet{jonsson2022CrackingCode, vandervlist2008TeachingMachine}). These technologies are still rarely integrated in design education \cite{dove2017UXDesign}, and often approached with the believe that even just exposing students to cutting edge technology can stimulate the emergence of innovative and technologically advanced design solutions to real problems \cite{mccardle2002ChallengeIntegrating}. In other cases, however, the approach is diametrically different: AI and ML are conceptualised as phenomena to be understood and questioned, because of its potential impact in society. For instance, \citet{auger2014LivingRobots} used the theoretical lens of domestication for challenging students to ideate future domestic robots and reflect on their implications in everyday settings.
The two perspectives on AI and ML, the focus of which we summarize as technical competency vs critique, tend to remain distinct approaches with apparently opposite scopes. Even when there is an explicit commitment to bridge the two approaches, existing pedagogy struggles to combine the ambition to build AI literacy while also fostering a critical mindset around AI/ML projects, and reflections do not lead to rich critiques about situated and contextual implications of AI and ML unless they are integrated into project development. For some counterexamples, 
\citet{jonsson2022CrackingCode} purposefully crafted a course for students to approach and appreciate AI tools as creative partners, and learned that AI qualities, such as uncertainty, imperfection, and under-determination, can be a rich source of inspiration for generating creative expressions as well as powerful triggers of reflection. 
Mital's Cultural Appropriation with Deep Learning course \cite{mital2021CulturalAppropriation}  weaves together learning about the operation of deep networks with recognising their role in society. Fiebrink's  work  \citet{fiebrink2019MachineLearning} distinctively looks at ML as a design material and situates it within project development. Perhaps most similar to the work outlined here is 'Graspable AI' \cite{ghajargar2021ExplainableAI, ghajargar2022GraspableAI, ghajargar2023MakingAI}, which brings together tangibility and AI, using explanation as a path to understanding and form as a language for communicating AI affordances. Even in these cases, however, the emphasis is on one side of the spectrum, that is on how to teach ML effectively to any population and enable the emergence of new creative outputs. 
%Considerations of how AI brings about distinct new challenges are also somewhat present in the author work, such as in their attempt to make up for the students’ natural tendency of thinking of computers as anthropomorphic entities– as hidden minds. However, these hardly account for the complex socio-technical implications of these technologies. 

The disciplinary call for exposing the design questions involved in making AI and ML systems -- as well as the complexity and trade-offs that implementing these in the world implies \cite{bilstrup2022DemoDesign} -- remains largely unanswered. 
Our work sits at the intersections of these experiences and aims to fill the gap between technical efforts and critical explorations. Specifically, we set out to integrate AI and ML explorations within the development of design projects, in a way that both enable students to build AI literacy, as well as to empower them to take a critical stand towards these technologies in society. %\todo{Ghajargar et. al. look at tangible approaches here, both for AI in general, and specifically for XAI; we differ, in that we are using a broader set of experiential methods, and ...}.

\subsection{Summary and research direction}
Part of the work of design as a discipline is to mediate between these philosophies and actionable practices that can be brought to bear on particular situations. That is the starting point for the work presented in this paper: we are interested in how to bring  conceptual developments from design theory and AI into something that is at hand to design students, that can make a difference to how they go about conceptualising and prototyping interactions. 

In order to bridge the gap between the practical and technical engagement with AI, we propose that three levels of engagement between AI and design are all potentially at play within design projects creating AI powered systems:
\begin{description}
    \item[\textbf{Interactional Affordances of AI}] that allow new means of interaction between systems and people. At a low level, AI brings new possibilities for sensing, responding, recognising and classifying from which to build interactions. These interactional affordances and possibilities for action \cite{stoffregen2003AffordancesProperties} offered by machine intelligence can take the form of capabilities offered by the technology (see \citet{yildirim2023CreatingDesign} for a comprehensive overview), but also of modulations of existing capabilities with AI specific qualities such as probabilistic outcomes. 
    \item[\textbf{AI Relationality}] as it is brought into constellations and forms new relations between people and things. Beyond the immediate interaction, design with AI intervenes conceptually and materially in constellations \cite{coulton2019MoreThanHuman} of humans and objects. Designers must navigate the increased agency and depth of interaction that intelligent systems bring, and the changes in the way that we understand and relate to technological systems.
    \item[\textbf{Wider Implications of AI}:] as it affects social structures and people’s lives outside the immediate interactions. Concerns about the implications of systems are not new, but AI and data driven systems that are built through processing large amounts of data about people bring new and subtle ways in which they can be unfair or unjust, more blurring of responsibility, and more potential unintended consequences at scale.
\end{description}

For the work at hand, we are interested in how these levels relate to design education, in particular how students start to engage with AI as a design material. To create a broad coverage, we looked at creating methods that could create engagement with the specifics of working with AI systems on these three levels, as well as balancing the educational concerns of developing a better understanding of and facility with the technology, and encouraging critique. Based on this, we created a set of ‘design exercises’ - rapid, experiential engagements that draw on the theoretical developments above but can be carried out productively in the context of conceptual development and prototyping interactions with AI systems (Figure \ref{fig:method-space}).

\fig{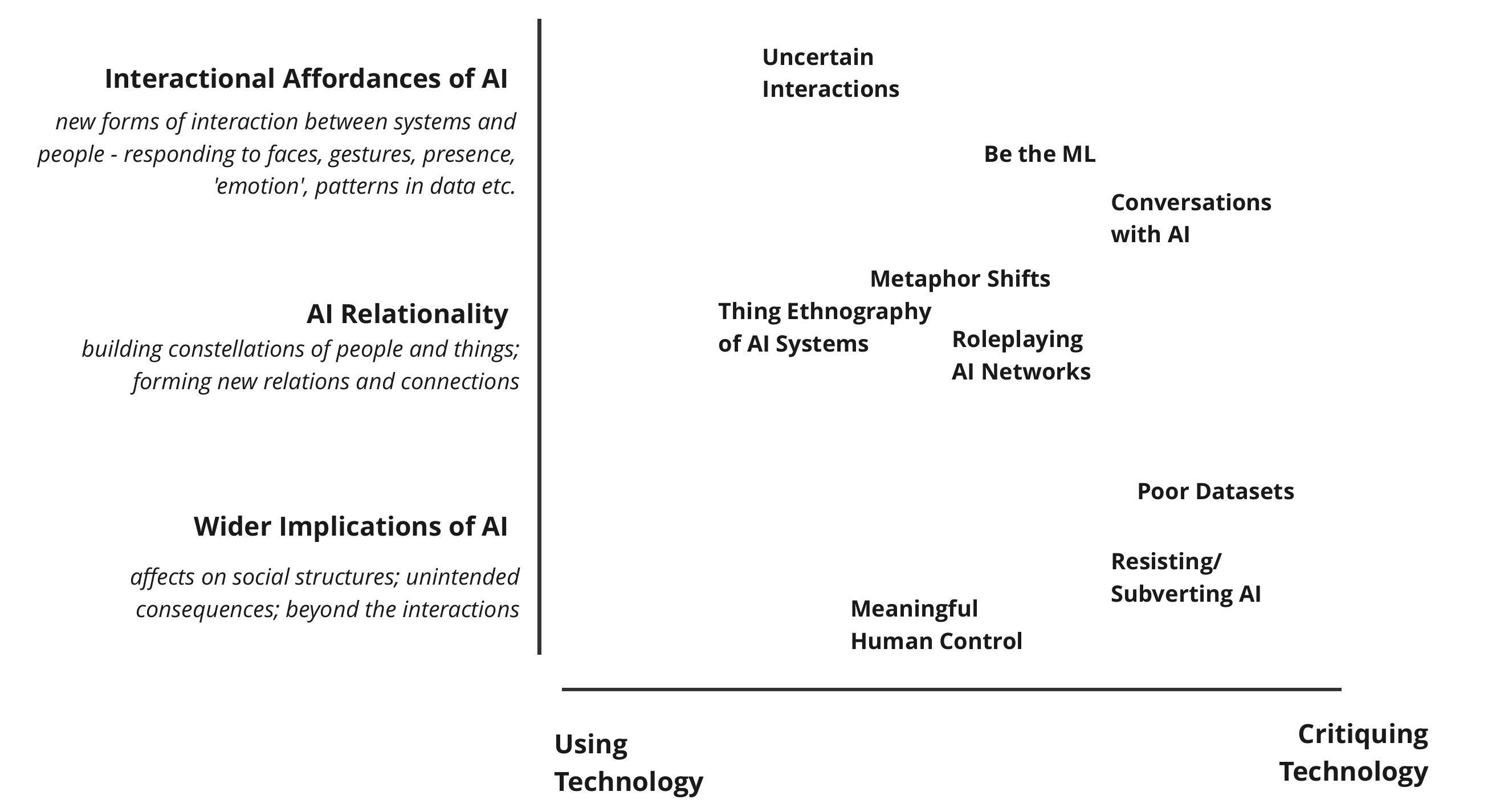}{method-space}{Situation of the methods underdiscussion across two axes: i) the level of consideration, from direct interactional afforances, through building relationality out to wider implications and ii) the balance between developing facility with the technology and critiquing it's uses.}

%% file: 03-Materials.tex
\section{Study}
\subsection{Course and context}
\label{sec:course-context}
The context of this study is a one-semester (20 week) design and prototyping course for first year Masters students in the `Design for Interaction'  programme at TU Delft. All students in the course (n=100) have a design background with a mixed range of computational skills, from no technical knowledge to beginner level in software engineering. The students were grouped by the course coordinators into 28 teams, and coached by 7 experienced coaches from the Industrial Design Engineering faculty.

The course is structured in three stages, with student-teams of 3 or 4 students working 13 per week on their design project (Figure \ref{fig:COURSE}). They worked on design briefs that asked them to speculate about near future interactions supported by technology. Many of these briefs were provided by client companies, for example new forms of human-vehicle interactions, possibilities for more sustainable cooking through smart kitchen appliances and pervasive computing in hotel rooms. 
The students had little to no pre-course familiarity with Machine Learning and AI methods, theories and tools, however most of the coaches had at least some experience with these technologies.
%The lecturers had competent to proficient levels of expertise and familiarity with AI and ML, either through real world design projects or academic research projects. 

The main learning objective of the course is to introduce students to various ways of prototyping with interactive technology. Students were asked to design within the context of the client company, creating and testing a new iteration of their prototype each week in discussion with their coach. They were prompted to draw on some form of AI or ML, although technical capabilities could be acted out rather than implemented in code. 
%Students were provided with a tech kit, containing an Arduino Sense BLE and a limited set of sensors and actuators as a starting point. Weekly coach sessions were used to discuss the progress in the student teams’ design projects, both conceptually and with regard to their prototypes. %their understanding of AI, Machine learning and its applicability in the teams’ design briefs as well as their level of prototyping with interactive technology. 
The course ran in three phases: a ``First Shot'' familiarisation with AI and technology, ``Iterating Forward'' to develop concepts and ``Polishing Up'' the final ideas and prototypes (Figure \ref{fig:COURSE}), with an exhibition at the end of each stage. Client companies were invited at the end of each stage to provide feedback to the student teams, organised in the form of an exhibition with interactive prototypes. 

\fig[1.0]{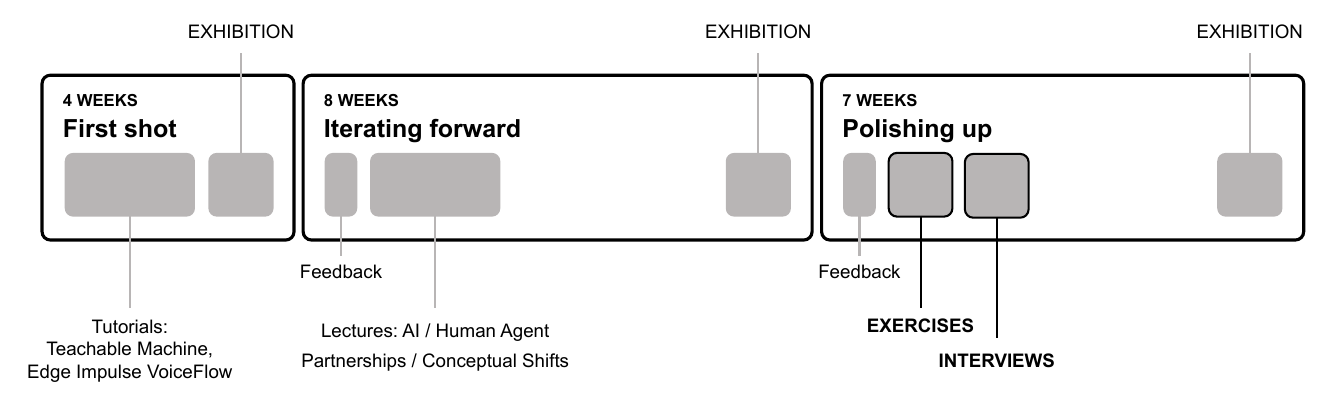}{COURSE}{Course Structure; In the first stage (4 weeks) students were given context on AI and ML, and hands-on engagements with AI technology were provided through a series of workshops with existing tools (Edge Impulse, Teachable Machine and Voiceflow). At the end of the first stage, teams presented multiple ideas demonstrated in multiple early prototypes. The second stage introduced lectures covering AI capabilities, Human-Agent partnerships and the conceptual shifts mentioned above as the students developed their core concept, leading to a second exhibition of interactive prototypes. The third stage introduced the exercises discussed in this paper, as the students refined their projects towards a highly immersive final exhibition with one or more interactive prototypes.}

\subsection{Exercises}
\label{sec:methods}
The intervention involved a set of 9 exercises (Table \ref{table:METHODS}).  Each design exercise was introduced on a single page, containing a title, a short description and instructions on how to execute the exercise, a background section describing the intent, usefulness and ideas behind the exercise and  references to papers and related projects (Figure \ref{fig:EXAMPLE-CARD}, see supplementary material for full set). The choice of this set of exercises was exploratory: we derived them from a combination of existing design practices, emerging work from the researchers and the theories mentioned above, through extensive discussion between the researchers. We aimed to have a spread of exercises across immediate interactional affordances of AI, mid-level human-machine relations and concerns about the wider implications of AI, as well as across developing fluency and supporting critique (Figure \ref{fig:method-space}, description in Table \ref{table:METHODS}, details in Supplementary Material). Some of the methods were pre-existing explorations, some had been used extensively, and some were adaptations of existing techniques to fit the AI context or the autonomous format. There was  a strong focus on activities that could be performed relatively simply by students, that were experiential, and that would work across a range of topics and levels of technical accomplishment. Each exercise was intended for application to an existing project, i.e. not a brainstorming or early ideation tool, but a way to develop existing work.

\fig[1.0]{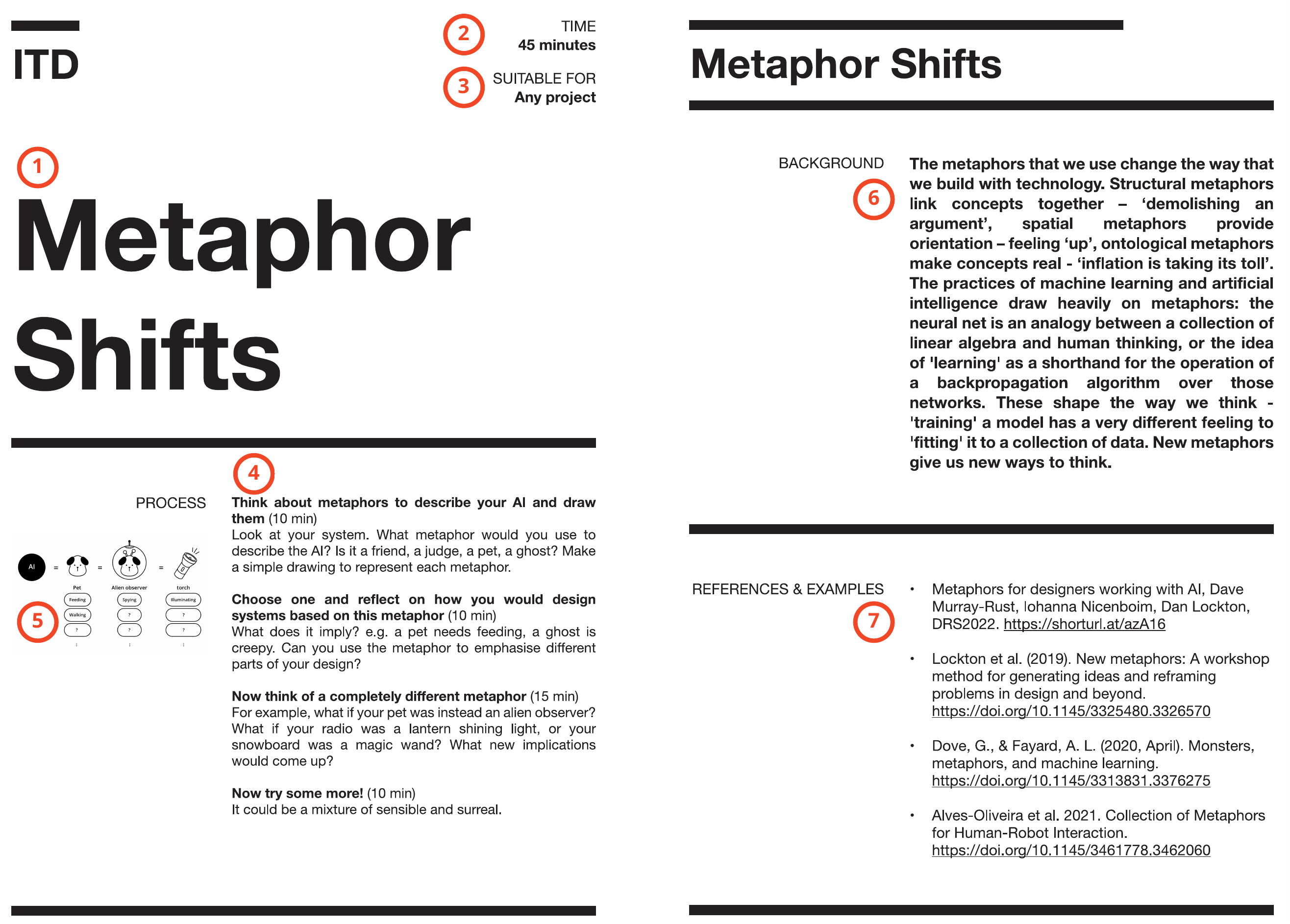}{EXAMPLE-CARD}{Example of the exercises, showing 1) title, 2) expected time 3) suitable project types 4) process 5) custom illustration 6) background, 7) references and example projects (full set of exercises in Supplementary Material)}

\input{figures/method_descriptions.tex}

\subsection{Execution and Data Collection}
Towards the start of the third stage, after feedback from the second exhibition (Figure \ref{fig:COURSE}), a half-day workshop was organised in which all student teams were introduced to the 9 design exercises with the aim of refining their project. This timing was chosen primarily for educational reasons -- the methods here were designed to help develop and refine existing ideas, rather than generate new ones, so we waited until the final stage of the course. This timing was the subject of some discussion -- see Section \ref{sec:timing}

This half day workshop allowed each group to execute one or two of the design exercises for their design project. Output of the half-day workshop was captured on A3 templates, including a questionnaire with some first prompts on the usefulness and effectiveness of the exercises applied. The half-day workshop was setup to be executed autonomously by the student teams, selecting the exercises themselves, with lecturers present to observe and assist when necessary. All output materials of the design exercises during the workshops were collected afterwards. 

Two weeks after the workshop we invited each team to select a representative to take part in a one-to-one semi-structured interview to discuss their experience and the effect it had on their project. In order to minimise educational disruption at a busy time and limit the possibility of coercion, we did not attempt to get full coverage, but allowed self selection by the students, in return for a €20 contribution to the teams prototyping budget. This led to 12 out of 28 teams participating in the interview. %To minimise educational disruption, we interviewed a subset of the teams ($n=12$ from $28$), selected to give a range of levels of engagement with theory as judged by coaches as well as from a first analysis of the teams’ output materials of the design exercises on the half-day workshop. 
We interviewed all coaches the week before the end of the course ($n=7$) to see what effect they had perceived on the students work. Interview questions and structure can be found in the Supplementary Material.

\subsection{Analysis and evaluation}
The interviews were audio recorded and both the interviews as well as the output materials from the design exercises were transcribed and analysed by a team of 7 researchers. We inductively coded both the written materials and interviews with students and interviews with coaches. We conducted collaborative thematic analysis: the coding team collectively familiarized themselves with the data and defined a shared coding scheme. At least two members of the team coded each of the transcribed materials using this scheme. Finally, coded materials were collectively discussed to synthesize insights into key themes, framed by the three levels of engagement with AI discussed earlier.

%% file: figures/method_descriptions.tex
\begin{table}
\begin{center}
\caption{Name, key references and description of each method. The references given here are for the theoretical context and inspiration for the methods, and do not match exactly those given on the exercise cards which prioritise developing student understanding.}
\small
\begin{tabular}{p{2.2cm}p{8.5cm}}
\toprule
Method & Description and inspirations \\
\hline
Uncertain Inter\-actions & Look through the state diagram of your interaction; for each change of state, imagine replacing it with probabilistic, uncertain or in-between outcomes \cite{bowler2022ExploringUncertainty, benjamin2021MachineLearning} \\
\hline
Be the ML & From a live view of the data inputs to your system, try to perform the activity yourself, then explain what you’re doing to someone else \cite{devendorf2015BeingMachine, scurto2021PrototypingMachine} \\
\hline
Poor Datasets  & Iteratively remove examples from your dataset, decreasing diversity of the input and retraining the model until something problematic happens. \cite{elwes2019Zizi, buolamwini2018GenderShades} \\
\hline
Thing Ethno\-graphy of AI Systems &  Collect data from the perspective an object in your system; use it make sense of the situation around the device; what does it experience that you didn’t know? Who does it interact with? \cite{giaccardi2016ThingEthnography, murray-rust2019EntangledEthnography} \\
\hline
Conver\-sations with AI &  Choose one of the AI powered objects in your interaction. One team member plays the character of the object, and others carry out an interview with that object. \cite{nicenboim2020MoreThanHumanDesign, reddy2021MakingEveryday} \\
\hline
Metaphor Shifts & Think about the metaphors  use in describe your system, then think about designing purely for the metaphor; change metaphors and try again \cite{murray-rust2022MetaphorsDesigners, lockton2019NewMetaphors, alves-oliveira2021CollectionMetaphors} \\
\hline
Roleplaying AI\newline Networks &  Play an object, system or human role and collaboratively act out the interaction. Discover new actors, negotiations,  relationships and interaction details \cite{pschetz2019AutonomousDistributed, reddy2020EncounteringEthics} \\
\hline
Resisting/ Subverting AI &  Recognise and act out moments in the interaction where someone might subvert the interaction; look for design opportunities \cite{lupetti2020SubversiveCitizen, disalvo2015AdversarialDesign} \\
\hline
Meaningful Human Control & Brainstorm places where the interaction might go wrong; rather than trying to fix it, figure out who or what is responsible and how they might be supported in setting their moral boundaries. \cite{siebert2022MeaningfulHuman} \\
\bottomrule
\end{tabular}

\label{table:METHODS}
\end{center}
\end{table}

%% file: 04-Findings.tex
\section{Findings}
Our findings are structured in two parts which build on both the A3 worksheets ($n=28$) and the student and coach interviews ($n=12$, $n=7$ respectively). While the analysis of the A3 sheets revealed recurring topics and common themes, the interviews revealed in-depth insights about what the students took from the exercises. The first part (Section \ref{sec:method-execution}) covers the execution of the methods: which ones were chosen and how they were perceived and valued by the students. The second part (Section \ref{sec:ai-themes}) describes the links made to AI and machine learning at the interactional and relational levels as well as wider implications. In all cases, comments from student interviews are marked as \qtn{p\textlangle{}id\textrangle} and those from coaches as \qtn{C\textlangle{}id\textrangle}; extra context about the project that the quote relates to is given in square brackets. The students who participated in interviews were working on projects around: comfort and behavioural encouragement while driving, as well as behaviour modelling and matchmaking (for Ford); collection of data while surfing and intelligent ski clothes (for O'Neill); smart objects and energy manifestation in hotel rooms (for Citizen M Hotels); photography for reconstructive surgery (for Erasmus Medical Centre); and speculating on spirituality and life coaching with AI (for the DCODE project).

%Before the in-depth findings from the interview transcripts (Section \ref{sec:interviews})  we build context from data on the A3 worksheets (n=28) produced by each group during the workshop: 

%; secondly, we present a qualitative analysis of the content of the A3 sheets. 

%

% We found that the exercises helped students in conceptual development, refining interactions, reflecting on their concepts, and understanding AI.  

% We first unpack how exactly the exercises supported students in those areas and exemplify them with responses from students, and then draw out the broader themes for engaging with AI.

\subsection{Method Execution}
\label{sec:method-execution}

To build context about the way the exercises were carried out, we give a quantitative summary of students' opinions of  their project before the workshop, and their evaluation of the clarity of the exercises. We then look more qualitatively at two themes: students' sense of the relevance and overall evaluation of the methods, and an analysis of the ways in which they found the methods useful. Table \ref{table:USAGE} summarises the number of times each method was and provides  key quotes for their use in these four areas: concept development, detailing interactions, understanding AI and supporting reflection. 

\subsubsection{Quantitative self-assessment and Clarity of the Methods}
Analysing the worksheets, we looked into how many times each method was used as well as the perceived clarity of the instructions (Table \ref{table:USAGE}). 20 of 28 groups carried out two exercises, with the remaining 8 carrying out only one. Counting responses to questions about their projects where a value greater than 0 was given (Figure \ref{fig:PROJECT-RATINGS}), 16 groups (0.57) felt their project critically investigated technology; 12 (0.43) were solving real-world problems; 15 (0.54) made use of AI qualities; 22 (0.79) engaged with complex relationships and 16 (0.57) intended to consider the wider implications of their work. 

All of the methods were rated as clear (Likert scale $-3=$"very unclear", $3=$"very clear",  $m=1.5, sd=1.0, m_{min}=1.0$), with only a single instance being rated negatively. This indicates that students felt they understood the purpose and structure of each method.

\fig[1.0]{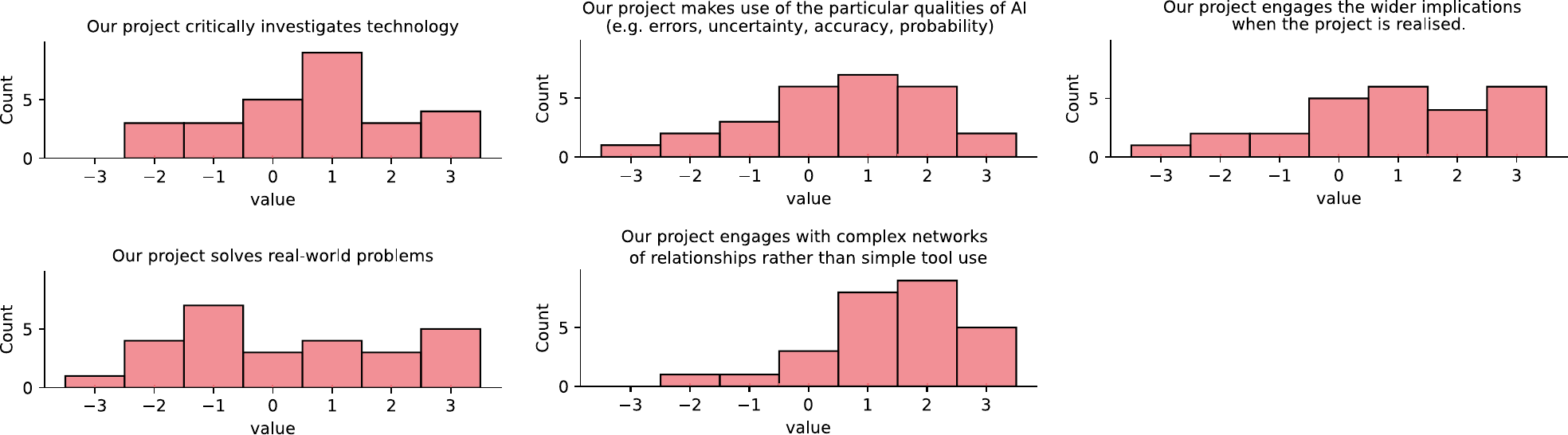}{PROJECT-RATINGS}{Student response (per group, n=28) to questions about their orientation and their project scope. Answers are on likert scales from ‘Not at all’ (-3) to ‘It’s the core of our project’ (+3).}

\input{figures/method_usage_short.tex}

\subsubsection{Relevance and situation in the course}
\label{sec:timing} 

Most groups chose exercises to address what they considered unexplored in their projects, or in some cases, even limitations of their concepts. For instance, Metaphor Shifts was picked to \qt{p1}{look for something else that better describes [their project]} or \qt{p15}{finding a nice metaphor [to make] interaction with the AI more empathetic to the user}. Roleplaying AI Networks offered the hope to be \qt{p27}{really precise and defined in the personality that we were gonna give the AI [that was deciding people's futures]}, and Uncertain Interactions was chosen to help \qt{p19}{map out all the responses and interactions that we were not considering before} in a multi-object interaction. Other groups saw the exercises as a more general way to \qt{p1}{check if we were in the right direction}, \qt{p6}{plan ahead like the possible problems}, \qt{p15}{have a discussion point [...] instead of just everybody thinking in different directions} or more radically \qt{p6}{just start again and we go somewhere else}. Some methods were explicitly avoided as the groups felt they had experienced the methods before, e.g. \qt{p19}{role playing}, or because they didn’t fully understand what a method entailed \qtn{p16}. Although the students reported that the methods were  relatively clear, they suggested that the individual differences might have impacted to some level how students interpreted the exercises: \qt{p1}{we're all from different cultures. So we all interpreted some questions differently}.

There was a common response that this activity would have been more useful earlier in the course  \qtn{p2,p5,p16,p19,p27, \ldots} and it could have helped generating more prototype ideas \qtn{p23}. Part of this was due to the sense that the activities felt like \qt{p2}{an ideation – like an inspiration activity}. As students saw the exercises as tools for divergent thinking, they would have liked to use them for ideation in close connection to the prototyping experimentations in the first period of the course \qtn{p5}. Others were  concerned that the moment when they did the exercises was their time to  \qt{p16}{optimize the prototype for the exhibition} and wanted to spend all of their time in making. In contrast, feedback from coaches on the timing was more positive: \qt{CIa}{To have sort of a zoomed out exercise at that point is I think a very powerful thing to do.[...] if you don't know where you're heading, then all these things, I don't think they will help you. [...] So I wouldn't move it. }. Several students echoed this perspective emphasising how it helped their process, e.g., \qt{p7}{because we were kind of stuck with our idea in general}, that it helped \qt{p20}{think of more details} around a developed idea.

Overall, there was a positive attitude  towards the activities, even from groups that were initially suspicious: \qt{p1}{we were quite surprised because we were thinking ‘Ohh workshop again. [...]  What's going [to come] out of it?’ [...] And then in the end there was actually some things that really helped us.}. Some negative responses (4)  revealed student’s concerns on  carrying out the exercise properly  \qtn{p5}, or spending too long on one interaction \qtn{p19}. Some  (2), had a hard time finding usefulness in the experience  \qtn{p15,p16} , as they were already familiar with the methods, as \qt{p15}{wouldn't say it brought me a new understanding because the metaphor is something we [already] had}.

\subsubsection{Perceived Utility of the Exercises}

Many students saw the exercises as a form of \qt{p2}{ideation, like an inspiration activity},\qt{p7}{ kind of a brainstorm} that can help \qt{p7}{to get a better idea}. Several groups noted that they came up with different and more interesting ideas \qtn{p5} and that they \qt{p23}{could use this new inspirations}. The methods were seen as useful for sharpening projects and defining practical next steps, such as planning for when things went wrong, or checklists of common concerns.
Metaphor Shifts was particularly generative of new ideas \qtn{p5,p16}. Beyond this, the methods were seen to help in the following four areas: 

\paragraph{Conceptual development:} The methods supported articulation and \qt{p2}{helped us get our story right, like the overall purpose of the concepts}, to develop \qt{p5}{a better detailed new metaphor} and to \qt{p7}{make [a] choice in what we wanted}. The benefit of gaining more conceptual clarity was also mentioned by some coaches \qtn{CLu}. Uncertain Interactions was seen as useful for mapping out the edges of a concept, so that students could easily get into details and next steps \qtn{p6}. The methods also helped grounding ideas, asking whether the concept \qt{p1}{can also work on the AI or do we need some future technologies that are not there yet to make it real}. In some specific instances, the exercise helped to \qt{p15}{start thinking about time}, or to \qt{p5}{find new ways of taking the same idea and spreading it}. The exercises helped to get an overview of things that students should think about \qtn{p1}, making implications concrete and graspabale, in a way that is \qt{p7}{so in your face that you don’t even think about the fact that it will be in the future}. 

\paragraph{Refining Interactions:} Many groups came out with a more refined idea about how their conceptual interaction should play out, as the exercise \qt{p6}{asks you to go into parts that maybe you don't want to explore} and make projects more well rounded. The activities also helped students define interaction contexts better. Groups felt invited to \qt{p15}{draw [AI] already in the context}, and to \qt{p19}{think about the interaction with some of the objects in the [interactive hotel room] scenario}. Refinements also pushed them to account for the potential meaningfulness of the projects, to \qt{p20}{clarify intentions}, and anticipate outcomes, e.g.,  \qt{p19}{what happens if the user doesn't understand what [the smart objects are] talking about}. The experiential nature of the exercises helped to \qt{p2}{translate something abstract as "being challenged" or "supporting" [good behaviour while driving] to something actually tangible}, to think into the \qt{p15}{aesthetic experience} of AI where the \qt{p15}{metaphor [of ritual cleansing for data collection] helped to think about materials as well}.

\paragraph{Reflection:} The workshops were seen as a moment of reflection, a break from the \qt{p25}{many layers in such a project} to focus on particular aspects. This could be on a technical level for the groups who \qt{p2}{never really took the time to think about AI} or more interactional when they \qt{p27}{stopped to think about this character sort of thing}. There was developing a \qt{p25}{critical lens, in terms of moral responsibility} and \qt{p6}{seeing how important this is, to acknowledge the mistakes, to be trustworthy}. Beyond the initial designerly sense of responsibility, they engaged with broader factors contributing to \qt{p25}{moral responsibility for an AI system [that encouraged spirituality]}. Overall, the moment for reflection was seen positively, developing aspects of their work that were not thought through, and a sense that \qt{p6}{confidence comes once you [...] manage the critical points} of the interactions.

\paragraph{Understanding AI:} The workshop improved the confidence of students about working with AI, as \qt{p2}{ before the course it was just like ‘I don't know how to use an algorithm to do something cool’ [...]  and this makes it kind of [makes] everything just specific in one one workshop}. This was often not based on a deeper technical understanding of algorithmic operation, but on a thinking about how the AI would relate to things around it. Some groups ended up \qt{p7}{actually using more AI because of this [workshop]}, with confidence coming from \qt{p6}{now that we know what's going wrong, and we know how to respond to that}.

\fig{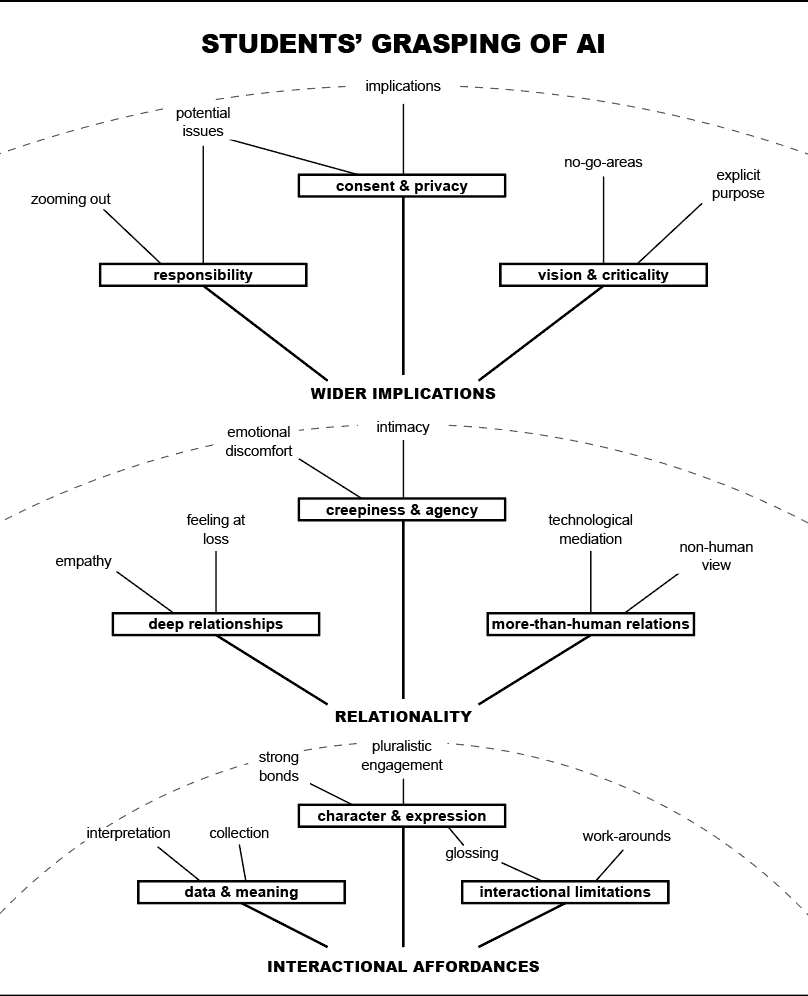}{concept-map}{Conceptual Map of students' reflections on the benefits of the methods grouped across the three levels of AI engagement: interactional affordances, relationality and wider implications.}

\subsection{Key themes for engaging with AI}
\label{sec:ai-themes}

Now we discuss findings in relation to broader theoretical developments in HCI, according to the three levels we have identified earlier: interactional affordances, relational questions and wider implications. An overview of these findings can be seen in Figure \ref{fig:concept-map}

%\subsection{Translating AI concepts for designers}
%In this section we analyse the findings in relation to the themes around the three levels of AI concerns developed earlier: interactional affordances, relational questions and wider implications.
\subsubsection{AI Interactional Affordances}
Students found that the  workshops illustrated that \qt{p7}{there are actually a lot of possibilities with AI}, beyond the tutorials at the start of the course, and that working through the experiences left them with a \qt{p2}{whole list of things that [AI] could say or do}. They already had some experience with particular topics, but this opened up a greater sense of how these possibilities could be deployed in relation to their work. This did not always change the concept of the interaction, but did provide a confidence that many interactional designs could potentially be realised.

\paragraph{Data and meaning:} Role playing helped with sensitising the students to the role of data in AI driven systems, questioning \qt{p27}{where is the AI getting the information?} both generally and through very detailed questions of \qt{p5}{where we’re gonna put [the camera that understands human-vehicle interactions]}. The coaches noticed the attention to physical detail as well, seeing development of a \qt{CIa}{way of bringing the data and looking at it and experiencing it}. There were moves to think about how to work with people in wheelchairs, and what it would mean to \qt{CLu}{recognise these things and build the dataset}, as well as the broader question of how \qt{CIa}{how [the collected data] can be meaningful for you as a person}.

\paragraph{Character and expression:}
The experiential nature of the exercises was, unsurprisingly, suited to engaging with designerly questions of the character and expression of the autonomous parts of the system. Students notice the possibility that they could be \qt{p27}{really precise and defined in the personality that we were gonna give the AI}, questioning default assumptions about how the system might respond. With conversational agents, it was noted that \qt{p19}{there is a lot of space between the yes or no}, but also that working probabilistically could smooth out interactional challenges, so that humans \qt{p6}{don't have to become machines ourselves}. The possibility arose to create pluralistic engagements that gave \qt{p23}{different answers based on different characters and based on different situations [for patients undergoing reconstructive surgery]}. This opened the possibility of making stronger bonds with users, and working on an emotional level, which we will return to in the next session.

\paragraph{Interactional Limitations:} In general, the coaches were more sensitive than the students to the potential limitations of technology, for example noticing when \qt{CGi}{the way they acted out looks good on screen but it doesn't reflect the deeper issues with understanding [...] Whereas if you use a conversational AI model I think you will run into a lot of problems that are hard to act out}. It was clear to them that some of the enactments would require sophisticated behaviour that could easily be glossed over with WoZ techniques, and they questioned whether the exercises could also point to these moments of glossing, or help notice points of complexity. For the more technically realised groups, the coaches noticed students working around limitations of the technology, where \qt{CGi}{ it was not very good at detecting facial expressions, but you made a hand gesture} that conveys emotion purposefully, leading to a rethinking of the interaction schema.

\subsubsection{AI relationality}
Students felt that \qt{p25}{[t]here are so many layers in such a project, where you are constantly building} and noted that the workshops took them into some of the complex, multilayered aspects of working with interactive AI systems. 

\paragraph{Deeper relationships:} Following the theme of character above, the workshops prompted students to think about the ways that humans related to the things being designed, giving an impetus to \qt{p15}{think more in an empathetic way} about the end users and what AI mediation would \qt{p25}{mean for a human to human relationships}. Roleplaying the situation with the device helped to look across some of the other people around the interaction, for example working with an system that was helping to take medical photos for reconstructive surgery  and seeing \qt{p23}{the relationships between the AI [and] doctor, assistant to friends or to your family members}. This was partly driven by a sense that the AI systems could interact in increasingly human-like ways, with metaphors like \qt{p5}{a friend in your car}, or a pet. There was a move to look at some of the longer-term relationships formed and the bonds that people made with AI systems. 
%One group became concerned they might \qt{p23}{make people feel at a loss after they need to give [their smart mirror] back to the hospital} after they had formed a bond with it. 
Students developed increasingly anthropic concerns from whether \qt{p23}{ people feel at a loss after they need to give [their smart mirror] back} at the end of a process, to questions of developing care and love relations to the objects.

\paragraph{Creepiness and agency:}
Interestingly, some of the more than human metaphors helped students to about when agency was troubling, and were \qt{p20}{open to more scenarios that we didn't see}. Manifesting home energy use using a metaphor of `fireflies' caused a concern that it \qt{p20}{will follow you through your room as a dog follows you. This might be kinda creepy. So what if they [users] don't want to be followed?}.
The potential intimacy of relations with a vehicle raised concerns about \qt{p5}{how intimate your interaction with your decentralized car be}, and how \qt{p5}{if you're driving and you're stressed and you somehow just get like this random unexpected hug from your car} it would cause emotional discomfort. Even when autonomous behaviour was not emotionally invasive, there were concerns that  \qt{p19}{sometimes the [smart hotel room] objects want to speak for themselves but at the same time you don't want to scare the human that is the guest in this room}.

\paragraph{More-than-human relations:}
Going beyond metaphors of caring for cars as one might care for a dog, coaches noticed that students would \qt{CIo}{use design as a medium to amplify the voice of nature} or \qt{CLu}{activate [...] energy consumption in a different way than just a tool} in their AI mediated interactions, making a shift to both non-human perspectives and the idea of technological mediation rather than tools for particular outcomes.
The students looked into new relationships that might emerge, e.g designing \qt{p16}{clothes to learn from every person that wears them to [and]  grow its own personality}.
The coaches noticed the roleplaying aspect of the exercises prompted critical reflection into the scenarios and relationships at hand, including noticing \qt{CGi}{that the setting that they were imagining and the role of the AI within that setting was not a very good fit}. Students found the practice useful for articulating what their vision for the future of human-AI relationships at individual and societal levels ought to be, including questions of governance and democracy.

\subsubsection{AI and wider implications}

\paragraph{Responsibility:} While methods targeted at interrogating control (Meaningful Human Control) explored  agency and control, other methods (Metaphor Shifts) still gave space for these questions to arise. Students reflected on \qt{p25}{considering moral responsibility for an AI system} within the creation process; and the coaches noticed that the workshops provided \qt{CIa}{a way to create distance and look at the project from a different perspective}, to re-evaluate the project beyond the immediate concerns of development, with a sense that  it was the designer’s responsibility to make sure that purposes and potential issues were clear upfront. Some students found that the workshops made the idea that people might misuse their system concrete, so for a friendly car system they \qt{p1}{gave ourselves some guidance for the next steps, [not]  for concepts [...], but more like OK, this is now a checklist that we need to put next through concept every time to make sure we think about this}. Responsibility often came through thinking through what might go wrong, with evidence of `zooming out' through the exercises, to think about what would happen if these systems were widespread, and their failure modes constantly present for users.

\paragraph{Consent and Privacy:} Several students mentioned issues around consent; while some felt this was a core part of their existing work, others found that discussion around the workshops was what they needed really understand the implications, and \qt{p1}{a solution for something that [is] difficult to think about}. Groups managed  to \qt{CLu}{dig deeper in that space} and better manifest the issues that they were already dealing with, and in some cases this meant that \qt{CGi}{[consent] was actually a very explicit part of their final concept and that was not at their departure, I think, was driven in part by going a bit more speculative than they were imagining at first}.

\paragraph{Vision and Criticality:} A common point from the students was that these workshops helped to think beyond the initial concerns of prototyping and into the multi-layered nature of the projects, not just the around AI responsibility but that \qt{p6}{it asks you to go into parts that maybe you don't want to explore} and rethink the purpose and shape of the project itself. Coaches were mixed about whether they saw changes in the level of critical thinking around the workshop, with some noticing no change, some a progression, and some seeing a strong difference where critical thought was brought in. Some of these were tradeoffs: \qt{CIo}{They became more critical. They were focussing more on the experience, but I'm not sure they were more engaged with the AI}. However, others noticed engaged with the human AI relationships, questions of datasets and the role of the project as critique,  and \qt{CMa}{really thought about it, how you negotiate with the machine and how much freedom you should have and how much agency you just have}.

%% file: figures/method_usage_short.tex
\begin{table}
\begin{center}
\caption{Name, key references and description of each method, along with the number of groups who used it and the average score for clarity. Yellow dots ({\color{myOrange}\CIRCLE\hspace{0.001cm}}) represent the number of groups who used the method. The final four columns address perceived usefulness of the exercise for four areas: concept development, refining interactions, understanding of AI and reflection. Red dots ({\color{myRed}\CIRCLE\hspace{0.001cm}}) count the number of groups who mentioned it as being helpful for the area, with representative direct quotes in italics, summaries in normal font.}
\small
\begin{tabular}{p{1.2cm}p{0.78cm}p{0.6cm}p{1.8cm}p{1.8cm}p{1.8cm}p{1.8cm}}
\toprule
Method & Clarity\newline {\scriptsize(-3..3)} & Count & Concept\newline Development & Detailing \newline Interactions & Understanding AI & Reflection \\
\hline
Uncertain Inter\-actions & 1.1 & \numUsesN{7} & \numUses{5} \qt{p24}{make it more concrete}, \qt{p6}{a screen to show why...}& \numUses{6} \qt{p7}{interaction in detail}, \qt{p24}{prepared for difficult conversations} & \numUses{2} \qt{p16}{better prepared for [] error interactions} & \numUses{2} \qt{p16}{reconsider intentions}\\
\hline
Be the ML & 2 & \numUsesN{1} & \numUses{1} \qt{p24}{gave us an overview}&  & \numUses{1} \qt{p24}{listening to human emotions is hard} &\\
\hline
Poor Datasets  & not used & not used  & & & & \\
\hline
Thing Ethno\-graphy of AI Systems & 2 & \numUsesN{2} & \numUses{2} \qt{p22}{the journey of the object outside the interaction} & \numUses{2} \qt{p22}{react to being held} & \numUses{1} \qt{p22}{[camera] viewpoint was not suitable} & \\
\hline
Conver\-sations with AI & 2.2 & \numUsesN{5} & \numUses{3} \qt{p2}{takes state of mind into account}, \qt{p3}{give AI a purpose} & \numUses{3} move from passive to active [p10], feel-good to good [p2] & \numUses{2} \qt{p3}{AI cannot have the same experience as you}, \qt{p2}{sensitizer} & \numUses{1} parts are separated, importance of roles (p10) \\
\hline
Metaphor Shifts & 1.8 & \numUsesN{14} & \numUses{11} \qt{p16}{role and character}, \qt{p7}{different perspectives} & \numUses{6} \qt{p3}{find a way to initiate negotiation}, \qt{p23}{gets a character}& \numUses{3}\qt{p18}{AI needs to be specific}, \qt{p3}{AI as a protective agent} & \numUses{5} \qt{p16}{how that intention has manifested itself into the concept} \\
\hline
Roleplaying AI\newline Networks & 1.2 & \numUsesN{5} & \numUses{4} \qt{p12}{new agents and interactions}, \qt{p13}{better clarity} & \numUses{4} \qt{p13}{draw out interactions}, \qt{p27}{elaborate on context} & \numUses{1} \qt{p13}{understanding of specific [technical] elements} & \numUseAlign none reported \\
\hline
Resisting/ Subverting AI & 1.0 & \numUsesN{3} & \numUses{3} \qt{p18}{exploring new ideas}, \qt{p8}{chances to be misused} & \numUses{3}  \qt{p28}{things we knew but never wrote down}, \qt{p8}{criticality}& \numUses{1} adversarial sensing (p8) & \numUses{1} \qt{p8}{think critically}\\
\hline
Meaningful Human Control & 1.0 & \numUsesN{7} & \numUses{5} \qt{p19}{making our design more responsible}, \qt{p2}{Introduce [...] resistance} & \numUses{5} \qt{p25}{development of consent}, \qt{p19}{part of interactions we haven't planned} & \numUses{3} \qt{p3}{understand the richness}, \qt{p2}{mess with your moral judgement} & \numUses{2}  \qt{p21}{consider user vulnerability}, \qt{p25}{bring across critical view} \\
\bottomrule
\end{tabular}

\label{table:USAGE}
\end{center}
\end{table}

%% file: 05-Discussion.tex
\section{Discussion}
In this discussion, we address some potentials for developing the exercises, and reflect on our initial research questions. We discuss how the AI exercises address the current methodological gaps and, more broadly, how this work contributes to a larger program around design, HCI and AI, nurture a distinctively designerly AI culture. 

\subsection{Effectiveness and Future Work}

The exercises were seen as effective overall, although they could further be improved through use, observation and iteration.  
They produced thoughtful, socially engaged responses, but to a large extent remained far from the rapid and technically grounded results generated at the beginning of the course, when students were provided with tutorials focused on learning a particular AI technologies. As an example, despite the deep technical grounding of Uncertain Interactions, most student responses did not get deep into the specifics of model output and how to make use of it. Future versions of the exercises could look to bridge this gap, as could their use in more technical contexts, were models were really being trained and deployed. There could also be support  to help students to decide which concerns to prioritise – for example, worries about people falling in love with their AI devices might not be the key problematics of the technology as created. While this prioritisation is arguably a part of general design practice, having concrete examples to contextualise the discoveries would be helpful. Practically, most students were relying on pre-built models and `Wizard of Oz' setups \cite{browne2019WizardOza, dahlback1993WizardOz} that used human action to simulate complex behaviour. This limited the utility of data driven exercises (e.g. Poor Datasets)  and fed into a focus on the anthropomorphic possibilities of AI. This also led to less engagement with the possibilities of new forms of human-machine interaction than we might have hoped for. 

\subsubsection{Timing and Situation}
The time that the students had to execute the exercises was short, which may have limited the potential for deep reflection and thoughtful practice. While the students still had access to the methods, few groups chose to make use of them, so there is space to explore more prolonged engagement. The positioning in the course was somewhat contentious, with many students feeling the methods had been introduced too late (Section \ref{sec:timing}) - this is coupled to their assumption that the methods were there for concept development and ideation. However, the overall feeling from the coaches was that the timing was sensible -- it provided a way to zoom out around existing concepts and add richness. Part of this divergence of opinion is part and parcel of process based education -- there are often different views from within the process than outside it. However, it does point to the need for a stronger sense of what one can expect from the methods, and an indication of when and how they could be productively deployed.

%Data collection was kept lightweight to avoid disrupting education. Since the students had not been given marks, there was some chance they were reluctant to deeply criticise the exercises.

\subsubsection{Choice and Range of Methods}

This initial set of exercises was based on a particular set of theoretical ideas; it is clear that other theories and concepts could prompt additional methods, and other methods could be derived from the theories used. There is certainly no shortage of candidates, whether agential cuts \cite{shotter2014AgentialRealism}  provide techniques to divide up complex systems and consider multiple boundaries through more or less embodied encounters \cite{vagg2022ExperiencingWithData}, or ideas of cyborg intentionality \cite{verbeek2008CyborgIntentionality} lead us to enact parings with composite possibilities \cite{rapp2021WearableTechnologies}, introspection provides a lens to think about relations between AI and lived experience \cite{brand2021DesignInquiry}. Methods with a clear technical genesis would offer immediate experiences that are deeply embedded in and shaped by the technology, for example deliberately misusing vision algorithms \cite{vanderburg2022CeciEst} or using computer vision as a site of enquiry \cite{malsattar2019DesigningPrototyping}. We see this as the start of a collection of ways to engage in this area, which will grow over time. Additional exercises might emphasize different parts of the design process and different modalities of experience as well as introducing new theories or grappling with particular qualities of AI.

\subsubsection{Applicability}

In terms of subject matter, the exercises were applicable to a range of projects across autonomous cars, robots, Internet of Things, hospitality and so on. They also helped with a range of issues, from shaping overall concepts to detailing important parts of the interaction. The application here was somewhat particular: the middle stages of an exploratory, creative prototyping brief. We would expect that the methods can be used in other processes and different levels of technical fidelity. In fact several of the methods, such as roleplaying AI networks and Thing Ethnography of AI systems are likely to give better results as the project is more developed and the context is stronger. Others, such as Poor Datasets are likely to be more useful with a  developed technical implementation, while Uncertain Interactions could help with ways to create interfaces around probabilistic models in deployment. 

%Aside from method choice, it is important to consider what made the exercises productive  for us as educators, and what is the benefit in making new ones? For us, there was value in creating the exercises, following the process of refining our thinking about the theories we were drawing on and developing clarity to articulate them to students. There was also something useful about having a mid-sized collection of complementary exercises: they 

%

\subsection{RQ1: Conceptualising and prototyping practices with AI}

The exercises illustrated some of the issues that students have when carrying out prototyping and conceptualisation with AI: the need to deal with uncertainty, the possibilities of more human-like interactions but less clearly defined capabilities, the need to hold multiple levels together. This clearly asks a lot from designers, especially in this case, where many of them did not have strong electronics and coding skills before the course. The experiential \cite{hemment2022ExperientialAI} and enacted \cite{elsden2017SpeculativeEnactments} aspects of the workshops were helpful to navigate this terrain, as the subjects of discussion could be played out in the group, adding to the sense of tangibility and refining how interactions should unfold. The interactional focus of this work makes it distinct from ideation tools such as AIxDesigns ideation cards \cite{AIIdeationCards} which focus on conceptual innovation, or work on developing user experiences \cite{subramonyam2021ProcessModel, subramonyam2021ProtoaiModelinformed} which makes the interface the primary subject of design. In line with open ended, critical and speculative prototyping methods \cite{malsattar2019DesigningPrototyping, vanderburg2022CeciEst, nicenboim2020MoreThanHumanDesign} the exercises took the students into the relational and interactional possibilities of AI.

From the feedback, it was important to give students exercises that were concrete enough that they could follow the steps. Several of the exercises were close to relatively standard design practices - Uncertain Interactions drew on the creation of state diagrams as a design articulation tool and the idea of acting out interactions as a form of prototyping is well established \cite{vanderhelm2020PrototypingTeam}. However, they were adapted to bring AI qualities into the familiar interaction design practices, emphasizing aspects like uncertainty, interface capabilities, distributed responsibilities and so on. It is clear that for some of the students, the simple forms of the exercises would have been enough – simply asking ‘what might go wrong’ and drawing a state machine to deal with it, rather than getting into the idea that machine learning systems produce probabilistic outputs produced useful results. None of the students chose to work directly with datasets; this may be a feature of their projects, as there was not much training and learning happening, or simply a lack of attraction to the particular exercise. %perhaps due to the ways that they were working, and that the amount of training and learning taking place was minimal.

There was a tendency with many of the groups to drift into anthropomorphisation, to imagine relations as overly human \cite{marenko2016AnimisticDesign}, and be diffuse about the capabilities of the technology. This relates to thinking into some of the particular AI characteristics that we will discuss in the next section, it is clear that prototyping will start to take different forms. The evolution of prompt engineering as a discipline \cite{liu2022DesignGuidelines} and the potential to generate working systems from prompts \cite[e.g. Aptly][]{2022IntroducingAptly} indicates that new forms of prototyping are emerging. Here, the constraints are less well defined than working with code on Arduino, but no less present - training a TeachableMachine \cite{carney2020TeachableMachine}  to detect a gesture has just as many concerns as using the electronic gesture sensor built into the Arduino BLE Sense the students were using, but the failure modes play out differently, and a different set of prototyping practices are brought to bear. 
The multiple viewpoints contained in the exercises here – people, things, datastreams, algorithms, networks – help to tease out the parts of interactions to prototype. Enacting these possibilities makes it easy to fall into broad, fuzzy, anthropomorphic thinking about what systems might do; the challenge for developing new forms of prototyping is to temper this with a grounding in the capacities of the systems being designed, and to engage with the human-like affordances of technology, without missing the new machinic possibilities. The exercises here helped students to clarify their concepts, move forward with their prototyping, and develop ideas about the responsibilities of creating AI systems,  while maintaining designerly concerns of materials, aesthetics, function, fit to context  and engagements with multiple actors.

\subsection{RQ2: Grasping  interactional, relational and contextual qualities of AI}

Our analysis of the student responses in relation to the current paradigm shifts in HCI, shows that interactional, relational and contextual qualities of AI could be important elements of design and AI educational programs. To unpack this, we look at our findings in relation with wider notions of agency, human-machine relations, and understandings of AI. 
\subsubsection{Agency}
As noted above, AI is a tricky term, but a lot is contained in the ideas of agency which it can develop, in particular around ‘non-humanesque agencies’ \cite{hildebrandt2020SmartTechnologies}. Some of these agencies were clear prompts for our work: a failure to recognise certain kinds of people as being human \cite{buolamwini2018GenderShades} shapes the inter-agences between vision systems and people; the ability to make decisions rapidly and constantly gives a sense of autonomy, but one which differs both in character and meaningfulness from the one of humans. Several of the exercises were aimed at interrogating these questions: ‘Be the AI’ prompted a reflection on exactly what the machines were doing, the interviewing and roleplaying exercises asked the participants to feel into what the agential possibilities were, and the more conceptual exercises questioned what agencies and responsibilities humans had around the systems. This helped participants to think about  \qt{p16}{what are the actual choices that we are gonna make or what part of the interaction are we gonna do ourselves and what part is the machine gonna do?} -- a key part of getting past the myths about AI capabilities \cite{natale2020ImaginingThinking}.

Much of the thinking reported around issues of agency had to do with \qt{p20}{how it's gonna be alive for people} - the clear, animate, characterful side of agency. This surely has roots in some of the roleplaying methods. The shape of the collective roleplaying exercises was informed by an increased emphasis on co-performance, as humans were brought in to act out what the smart technologies would do, and notice possibilities for more shared agency, whether co-learning with the AI or finding ways for objects to speak for themselves. 
\subsubsection{Human Machine Relations}
Some of the exercises prompted students to position AI in relation to humans and non-humans.  Thing Ethnography of AI systems instructed the students to map the ecosystem of the thing and its touchpoints, and reflect on who and what interacts with the concept. In Conversations with AI, the students were asked to enact an AI agent. Metaphor Shifts asked students to design systems based on a particular metaphor and then compare it to others. These exercises highlighted  the relations of humans and non-humans within AI systems \cite{coskun2022MorethanhumanConcepts, nicenboim2020MoreThanHumanDesign}. They did that by decentering the designers perspective, to consider more actors and interactions that go beyond one user and one device \cite{verbeek2020PoliticizingPostphenomenology}. They also invited students to relate to AI not only as a tool, but as a social agent that  shapes people’s lives. In students’ prototypes, intelligence, as well as responsibility, were not seen as a properties of machines alone, but shared between humans and artificial partners. Similarly,  uncertainty and unpredictability were  ‘collaboratively curated’ to ‘imagine forms of digital interaction’ \cite{marenko2016AnimisticDesign}.

The findings show that the exercises helped students to expand their concepts to account for the ecologies around them, especially when their projects were centred on particular embodiments, for example extending from a plant pot to a community of plants, and looking at the plant-plant relations as well as the plant-human ones.  The exercises also helped them acknowledge other people beyond the immediate users, thinking into how would they relate to the system, and what are the responsibilities that the user, system and designer have towards them. Furthermore, thinking of their concepts in relation to humans and non-humans, created awareness within the students on the human labour that is implicated in sustaining AI systems \cite{sinders2021LaborTools}. The kind of metaphorical social relationships the AI had with others ultimately influenced the designs: when the system was cast as a friend, it was seen, designed and conceptualised differently from when it was cast as a pet. 
\subsubsection{From Explanations to Understandings of AI}
One of the current challenges in the design of AI systems is how to support people in understanding them, especially when used to make autonomous decisions or create knowledge. AI explainability is especially challenging when based on deep learning models, given that some of the paths that AI systems use to give recommendations are not interpretable \cite{ehsan2020HumancenteredExplainable}, and the source of many generative outputs is complex \cite[e.g.][]{kovaleva2019RevealingDark}. While understanding ML in its technical sense is important, recent approaches in the explainability of AI have pointed at other ways of understandings which are not based on technical explanations and instead, promote experimentation, challenging boundaries, or promoting respect \cite{nicenboim2022ExplanationsShared,hemment2022ExperientialAI,seymour2022RespectLens}. The findings expand the agenda of Explainability of AI by illustrating and unpacking particular design engagements with AI that go beyond mastering ML technical capabilities. This points at particular aspects that are important in the kind of understandings that designers might need to gain of AI. Some particular aspects that helped designers understand AI are exercises that could prompt reflection into the affordances, relations and wider implications that those systems might have. Those engagements were not based on learning how to code ML models, but on experimenting with changing perspectives, provoking failures, enacting behaviours, and drawing schemas. These tactics could become part of a new agenda for supporting designers in understanding AI, especially one that is aligned with theoretical developments in HCI such as the posthuman turn \cite{lindgren2020SocialScience} as well as practical developments in design (such as methods used in critical, speculative and adversarial design \cite{disalvo2015AdversarialDesign,irani2014CriticalDesign, bozicyams2021PoeticsFuture}).

\subsection{RQ3: Critical design perspectives while engaging with AI as a socio-technical system}
While  the exercises helped the students to develop their projects (from ideation to conceptualisation and detailing) they especially illuminated and modulated changes in the students design processes in relation to the sociotechnical aspects of AI systems \cite{crawford2021AtlasAI}. The exercises supported students in reflecting on the role of AI within their concepts, in being more specific in what kind of aspects of AI are present, and developing a critical design perspective on AI around values of responsibility and agency.  

Designing with AI as a socio-technical system, means acknowledging that it is not only a technical domain, but also entangled with social practices, institutions and infrastructures, politics and culture. AI “is both embodied and material, made from natural resources, fuel, human labor, infrastructures, logistics, histories, and classifications”  \cite{crawford2021AtlasAI}. This is not an entirely new perspective – AI has long been considered a material practice \cite{agre1997ComputationHuman}, but there is a need to consider the interaction between humans and machines as part of broader societal contexts, and the broader discursive settings in which AI is socially constructed as a phenomenon with related hopes and fears  \cite{lindgren2020SocialScience}. 

From the findings, it seems the exercises provided a space for students to  go beyond the immediate concerns of a rapid prototyping session and engage in reflective practices that can position AI within its broader societal contexts. It is clear from many of the responses that the workshop carried out here provided a moment to reflect. Some of this can be simply ascribed to the sole fact of having the intervention -- a space that prompted further thought. However, some of the exercises were more specific triggers for reflective engagement, with Meaningful Human Control and Resisting/Subverting AI asking students to explicitly consider critical perspectives. There was evidence of ‘reflection-in-action’ \cite{yanow2009WhatReflectionInAction}, moments where in the midst of carrying out experiential exercises around the prototypes they ‘ma[de] previously implicit assumptions about the work explicit’ \cite{wegener2019ReflectioninActionWhen}. Where it had previously seemed that an insect swarm would give a warm sense of companionship, looking at the technological sense of surveillance and following revealed a darker possibility for the user; the idea that a supportive hug came from technical rather than human agency was found on reflection to be disturbing. 

Some of the exercises prompted a sense of ‘zooming out’ \cite{nicolini2009ZoomingZooming}, to consider wider networks of things and people, and this zooming out was part of the students move towards more empathic design. This touched on the temporal \cite{pschetz2018TemporalDesign} aspects of their design, as they thought not just about interactional moments, but the slower unfolding of relationships over time. Overall, there were many moves to develop a sense of criticality within their design: the AI oriented methods created viewpoints for considering the role of technology and its possible overreaches from pragmatic and whimsical perspective, in line with the ‘ongoing practical, critical, and generative acts of engagement’ \cite{suchman2020AgenciesTechnology} that build a responsibility for the things being designed.

\subsection{Beyond Education: Nurturing designerly AI cultures}
We started with the aim of helping students to grasp enough qualities of AI to adapt their processes and conceptualisations accordingly. The study highlighted that there are distinct ways to engage with AI that are appropriate to our setting, where the culture and practices of designers centre particular ways of working. In this section we discuss the aspects of our work that nurtures this designerly AI culture.

The use of AI technologies varies by field. If we look at AI in terms of the enabling technology and the culture that surrounds it \cite{caramiaux2022ExplorersUnknown}, some of the differences and parallels with the use of AI in design become clear. There are common moves to cast AI as a creative partner \cite{llano2022ExplainableComputational, mccormack2020DesignConsiderationsa} within music, as a solution for optimisation \cite{noor2017AIFuture} within engineering, as a formalisation and purification of human thought \cite{chiusi2020AutomatingSociety,singh2019DecisionProvenance} in decision making organisations and so on. Within design, the places that AI might sit are being negotiated. Do we bring it into the process as a sparring partner for ideation \cite{simeone2022PushingDivergence} or a source of creative inspiration \cite{yun2022IdeasquaresUtilizing}? Do we use it to re-understand the world through divergent practices \cite{malsattar2019DesigningPrototyping}? Is it a new computational capacity for which we have to develop new UX practices \cite{subramonyam2021ProcessModel}? Or a boundary object whose politics need critique \cite{crawford2019excavating,lyons2020ExcavatingExcavating}. All of these are within the remit of design. What we are interested in accenting here is the possibility for a designerly culture around the use of AI technologies, whether in processes, outcomes or critique. Just as a shift from explanation to shared understanding \cite{nicenboim2022ExplanationsShared} speaks to a relational, experiential mode of engagement, the exercises here create those experiences, and give ways to pick up, tangle and hold those relations. We suggest there are three key features of the methods that support this: experientiality, pragmatism and reflection.

The experiential nature of the methods appeared to be key in bringing in different perspectives on existing work, from noticing potential implications to uncovering new actors and interrogating positive ideas of agency. In this prototyping oriented style of working, enacting and dramatising possibilities helped to grasp concepts. This was particularly relevant to working with AI systems, where the level of agency expected of the technology is high, so vitalising it makes intuitive sense.

Secondly, the pragmatic nature of the exercises, distilling complex ideas down to a set of steps to explore supported critical discussion. Rather than starting from the theory, students were able to develop grounded experiences and respond to them. This led to practices such as developing their own checklists for responsibility as well as rethinking interactions based on new metaphors for the relations between technologies and humans.

Finally,  the exercises all point to building the skills that a reflective designer in AI might need – \qt{p16}{Perhaps the thing that they have in common is that they make you reconsider what your intention was and how that intention has manifested itself into the concept}. As such, they are distinct from technical support, even technical support tailored to creative practitioners \cite{2020AIxDesignCommunity}, but look to build bridges from  more than human thinking \cite{coskun2022MorethanhumanConcepts, coulton2019MoreThanHuman, giaccardi2020TechnologyMoreThanHuman, nicenboim2020MoreThanHumanDesign} towards technical practice. 

By providing this kind of multiple toolbox, we contribute to shaping the emerging AI-Design culture as something distinct from the technical, scientific, artistic and socio-legal cultures that are relatively well established. Further, we believe that this practice of grasping AI can be useful beyond the classroom, a powerful and versatile support for design professionals to meaningfully engage with the development of intelligent systems. 
\section{Conclusions}
There is a growing need for designers to engage with artificial intelligence and machine learning in their practice as it becomes integrated into the functioning of the physical and digital systems that they design. A particular challenge here is how to carry out ideation and early stage prototyping around AI/ML, when the exploratory nature of the work makes it impossible to invest much time in detailed technical understanding of particular algorithms or systems. At the same time, the technical possibilities of emerging algorithms can exert an overly large pull on designs, artificially narrowing the solution space and drawing away from the needs and qualities of the interaction. 

To develop the potential for designers to engage meaningfully in this space, working from an educational perspective, this paper introduced a series of ‘AI  exercises’ informed by recent theoretical developments in third wave HCI to help students grasp AI as a socio-technical system. We developed three levels of consideration for designing AI systems: interactional affordances, relational possibilities, and the wider social implications of AI systems; and provided methods for working at each level. Through qualitative analysis of these exercises with a group of students, we build up a picture of what kind of impact the interventions had on their understanding of AI and their project development. Through the exercises, the students refined their designs and clarified their concepts, and were able to move forwards with their prototyping with a greater sense of confidence in their designs and responsibility around the process. The experiential, pragmatic aspects of the exercises helped to make theoretical ideas concrete and generative of new possibilities, while keeping a sense of materiality and interaction with humans. The space for reflection provided by the exercises helped the students to develop a wider perspective on their work within the bounds of a rapid prototyping project.

The study findings highlight  ways in which experimental design exercises could support students in understanding AI, especially considering that such understanding needs to go beyond mastering ML technical qualities. The exercises here helped illuminate and modulate changes to the students design processes in relation to the interactional, relational and contextual qualities of AI, helping students develop a reflective and critical design perspective while responding to the key theoretical developments that are discussed in the AI community within HCI. Through the discussion, we raise questions of how a socio-technical view of AI, through ideas of agency and relationality can support a designerly culture around the development of AI.